\documentclass{ws-mpla}

\usepackage{graphicx}

\begin{document}

\markboth{Gabriella Sciolla}
{Directional detection of Dark Matter}

\catchline{}{}{}{}{}
\title{Directional detection of Dark Matter }

\author{\footnotesize Gabriella Sciolla}

\address{Massachusetts Institute of Technology, \\
Cambridge, MA 02421, USA\\
sciolla@mit.edu}

\maketitle

\pub{Received (November 2008)}{}

\begin{abstract}

Direct detection of WIMPs is key to understand the origin and composition of Dark Matter. Among the many experimental techniques available, those providing directional information have the potential of yielding an unambiguous observation of WIMPs even in the presence of insidious backgrounds. 
A measurement of the distribution of arrival direction of WIMPs can also discriminate between Galactic Dark Matter halo models.  
In this article, I will discuss the motivation for directional detectors and 
review the experimental techniques used by the  various experiments. 
I will then describe the DMTPC detector in more detail.

\end{abstract}

\ccode{PACS Nos.: include PACS Nos.}

\section{Introduction} 

 Understanding the nature of Dark Matter (DM) is one of the most fascinating and challenging goals of modern physics. Presently, our knowledge on Dark Matter comes entirely from astronomical and cosmological observations\cite{trimble}. 
We know that its existence is needed to explain the dynamics of galaxies\cite{Rubin} and  
galaxy clusters\cite{Zwicky}, and its distribution can be mapped using gravitational lensing\cite{lensing}. 
Recent data from WMAP\cite{wmap5years} indicate that Dark Matter is responsible for 23\% of the energy budget of the Universe and 83\% of its mass. 
Unfortunately, very little is known about the identity of Dark Matter and its interactions, because DM particles have never been directly observed in the laboratory. 
The direct detection of Dark Matter is, therefore, of utmost importance for understanding the dominant matter constituent of our Universe. 
  
 The most promising candidates for Dark Matter are axions\cite{axions} and Weakly Interacting Massive Particles (WIMPs). WIMPs are especially compelling in that a calculation of their relic abundance from production during the Big Bang is very close to the measured abundance if the particles are weakly interacting and have mass of the order of a few hundred GeV, which is approximately the weak scale. WIMPs arise in many extensions of the Standard Model of particle physics. In Supersymmetric models that conserve R-parity, for example, the Lightest Supersymmetric Particle must be stable, making it a natural candidate for Dark Matter. The favored candidate is the lightest neutralino, which is a mixture of the spin-1/2 partners of the photon, Z$^0$, and Higgs particles.

WIMP direct detection experiments look for 
nuclear recoils due to elastic scattering of DM particles on the nuclei in the active volume of the detector. A WIMP can scatter elastically from a nucleus 
 via two different interactions: spin-independent (SI) and spin-dependent (SD). For spin-independent  interactions, 
the WIMP-nucleus cross-section simply scales with the square of the mass of the nucleus, favoring target materials with high atomic mass. 
Spin-dependent interactions,  
in which the spin of the WIMP couples to the 
spin of the nucleus,  require  target materials with  nonzero spin nuclei\cite{Jungman} (see Table~\ref{tableF}). 
Spin-dependent cross sections are predicted to dominate over SI cross-sections in models
in which the lightest super-symmetric particle has a substantial Higgsino contribution\cite{SDTheory} .

\begin{figure*}[t]
\centering
\includegraphics[width=2.3in]{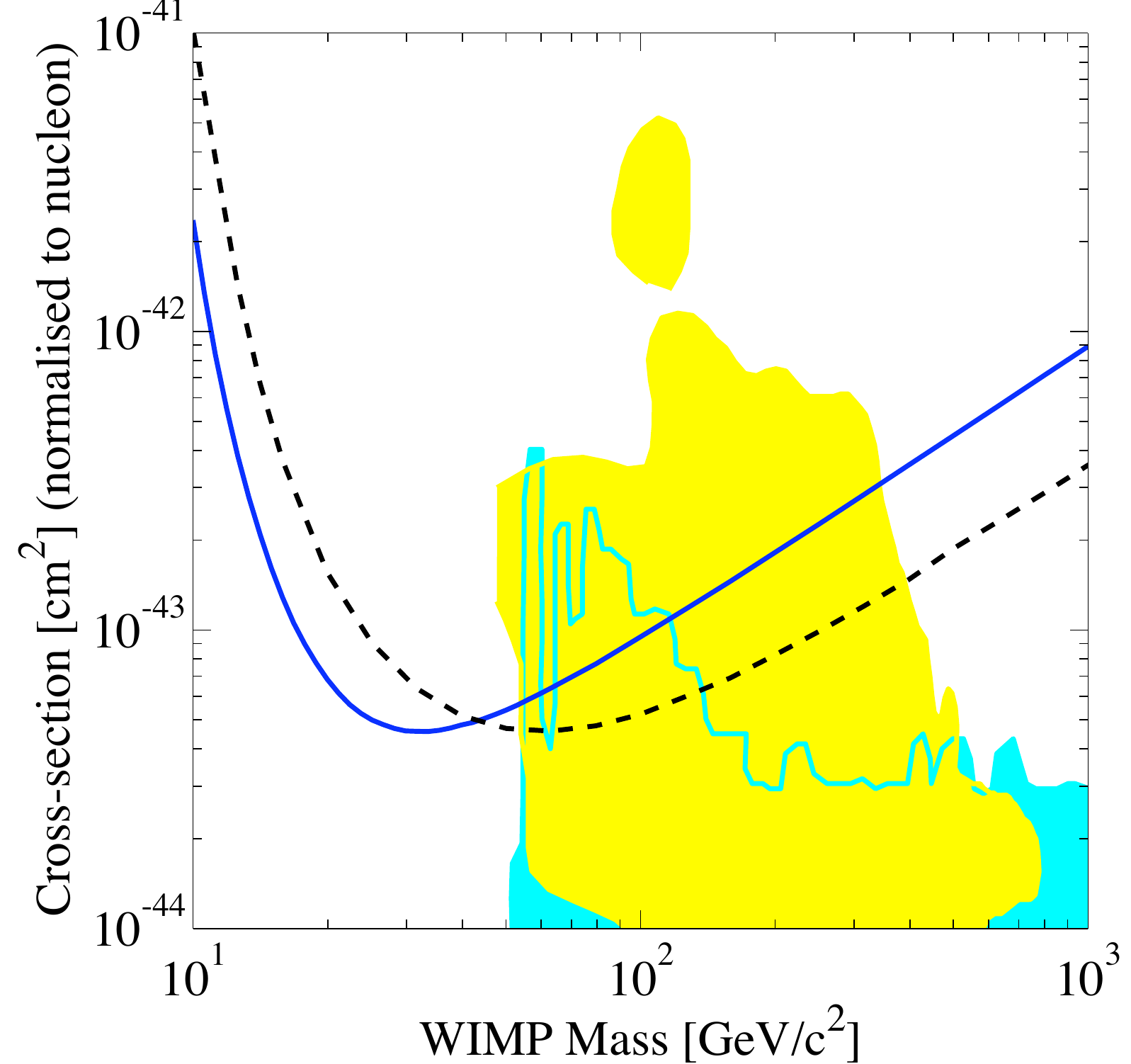}
\includegraphics[width=2.3in]{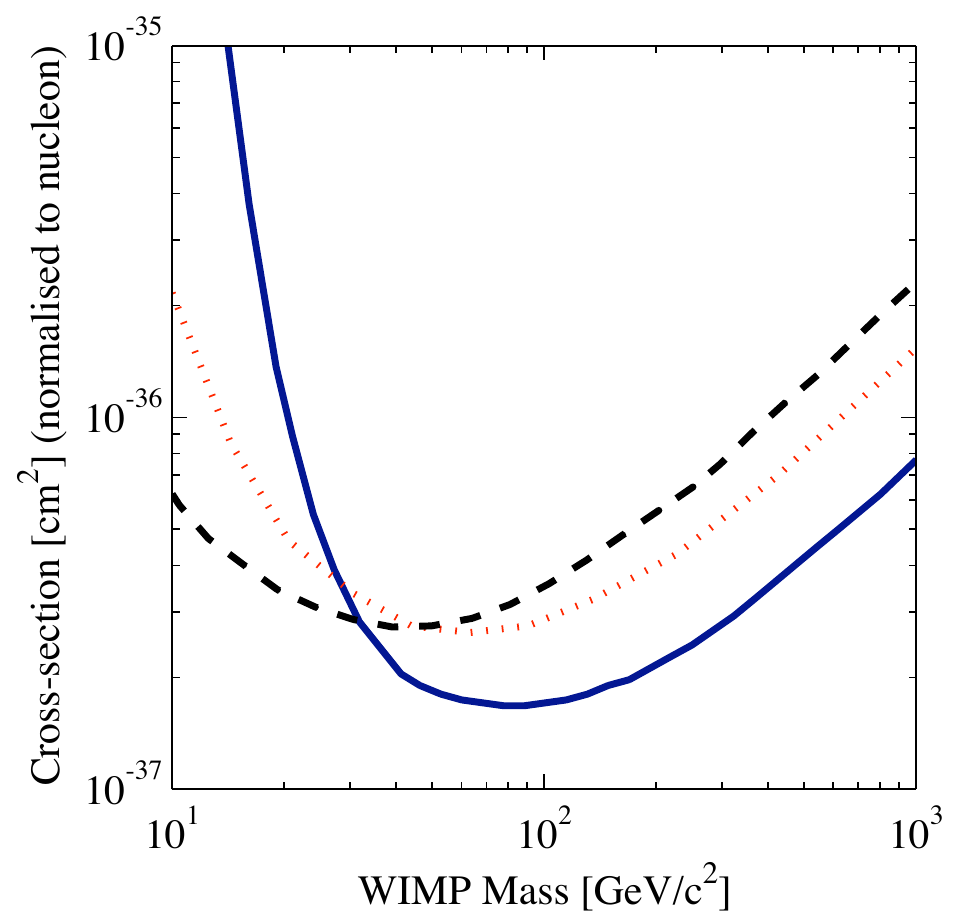}
\caption[]{
Recent cross-section limits as a function of the WIMP mass.\cite{filippini}  
Left: limits on spin-independent interactions from  
CDMS\cite{CDMS2008-SI} (solid line) and XENON\cite{XENON2007-SI} (dashed line). 
The yellow and blue areas show allowed region from theory.\cite{Baltz:2004aw,Roszkowski:2007fd}
Right: limits on spin-dependent interactions  on protons from  KIMS\cite{KIMS2007} (solid line), COUPP\cite{COUPP2008} (dashed line), and NAIAD\cite{NAIAD2005} (dotted line). 
}
\label{limits}
\end{figure*}

 Many experiments have been built to detect WIMPs in the laboratory using different experimental techniques. 
Figure~\ref{limits} shows the leading  experimental limits\cite{filippini} 
on the spin-independent (left) and spin-dependent (right) cross-sections as a function of the WIMP mass. 
While the limits on SI interactions are already cutting into the theoretically favored region, 
those on SD interactions, which are presently  seven orders of magnitude weaker than the corresponding  SI limits, 
are still  two orders of magnitude above the theoretical predictions.  
A substantial experimental effort is therefore needed to better explore SD interactions. 

One of the main challenges in direct detection of Dark Matter is suppression of backgrounds that mimic WIMP-induced nuclear recoils. 
Today's leading experiments have achieved excellent rejection of the backgrounds that have a distinct signature in the 
detector, i.e. photons, electrons and alpha particles.  
Unfortunately, there are sources of backgrounds for which the detector response is nearly 
identical to that of a WIMP-induced recoil, such as 
elastic scattering of neutrons produced either by natural radioactivity or by high-energy cosmic rays.  
Neutron backgrounds are presently reduced below detection threshold by 
making use of radio-pure materials in the active area of the experiment, by utilizing  active and passive shielding, 
and by locating the experiments in deep underground caverns. 
The absence of expected backgrounds simplifies  the data analysis, since all observed events are interpreted as signal. 
However, as the scale of DM experiments  grows to fiducial masses of several tons, 
it will be very hard to completely suppress all neutron backgrounds. 
In addition, ton-scale experiments  will start to detect a 
new source of irreducible background due to 
the coherent scattering of solar neutrinos,\cite{monroe} which no amount of shielding can reduce. 


The presence of 
neutrons and neutrinos will substantially complicate the interpretation of the data, 
as these insidious backgrounds are not only impossible to reject, but also notoriously difficult to predict.\cite{MeiHime} 
Therefore, when the first hint of a signal of dark matter will be detected, it may be difficult to convince the scientific 
community about the soundness of the discovery. 

A precise knowledge of all backgrounds would not be needed to produce an unambiguous observation of Dark Matter if one 
could correlate the observation of a nuclear recoil in the detector with some unique astrophysical signature which no 
background could mimic. This is the idea that motivates directional detection of Dark Matter.

\section{The need for directional detection of Dark Matter }

The observed  rotation curve of our Galaxy suggests that at the galactic radius of the Sun, the galactic potential 
has a significant contribution from Dark Matter. 
The Dark Matter distribution in our Galaxy, however,  is poorly constrained. 
A commonly used DM distribution, the standard dark halo model,\cite{isothermal} 
assumes an isothermal sphere extending up to 50 kpc from the galactic  center. 
The ordinary matter, which forms 
a disk-like luminous spiral structure at the center, rotates with respect to the halo. 
Observations show that our solar system orbits around the galactic center on a nearly circular path 
with an average velocity of about 220 km/s with respect to the DM halo. Therefore, an observer on Earth 
would see a wind of WIMP particles with average velocity of 220 km/s. In this model, the DM velocity is described by a Maxwell-Boltzman distribution with dispersion $\sigma_v=155$ km/s. 

The Earth's motion relative to the galactic halo leads to an annual modulation\cite{drukier}  of the rates 
of interactions observed above a certain threshold in direct detection experiments. 
In its seasonal rotation around the Sun, the Earth's velocity has a component that is  anti-parallel to the DM wind during the summer, and parallel to it during the winter. Since the velocity of the Earth is about 30 km/s, 
the apparent velocity of the DM wind will increase (decrease) by about 10\% in summer (winter), 
leading to a corresponding increase (decrease) of the observed rates in DM detectors. 
Unfortunately, this effect is difficult to detect because the seasonal modulation is expected to be small (a few \%) 
and very hard to disentangle from other systematic effects, such as the seasonal dependence of background rates. 
These experimental difficulties cast a shadow on the recent claimed observation of the yearly asymmetry by the DAMA/LIBRA 
collaboration.\cite{damalibra}

\begin{figure*}[t]
\centering
\includegraphics[width=2.5in]{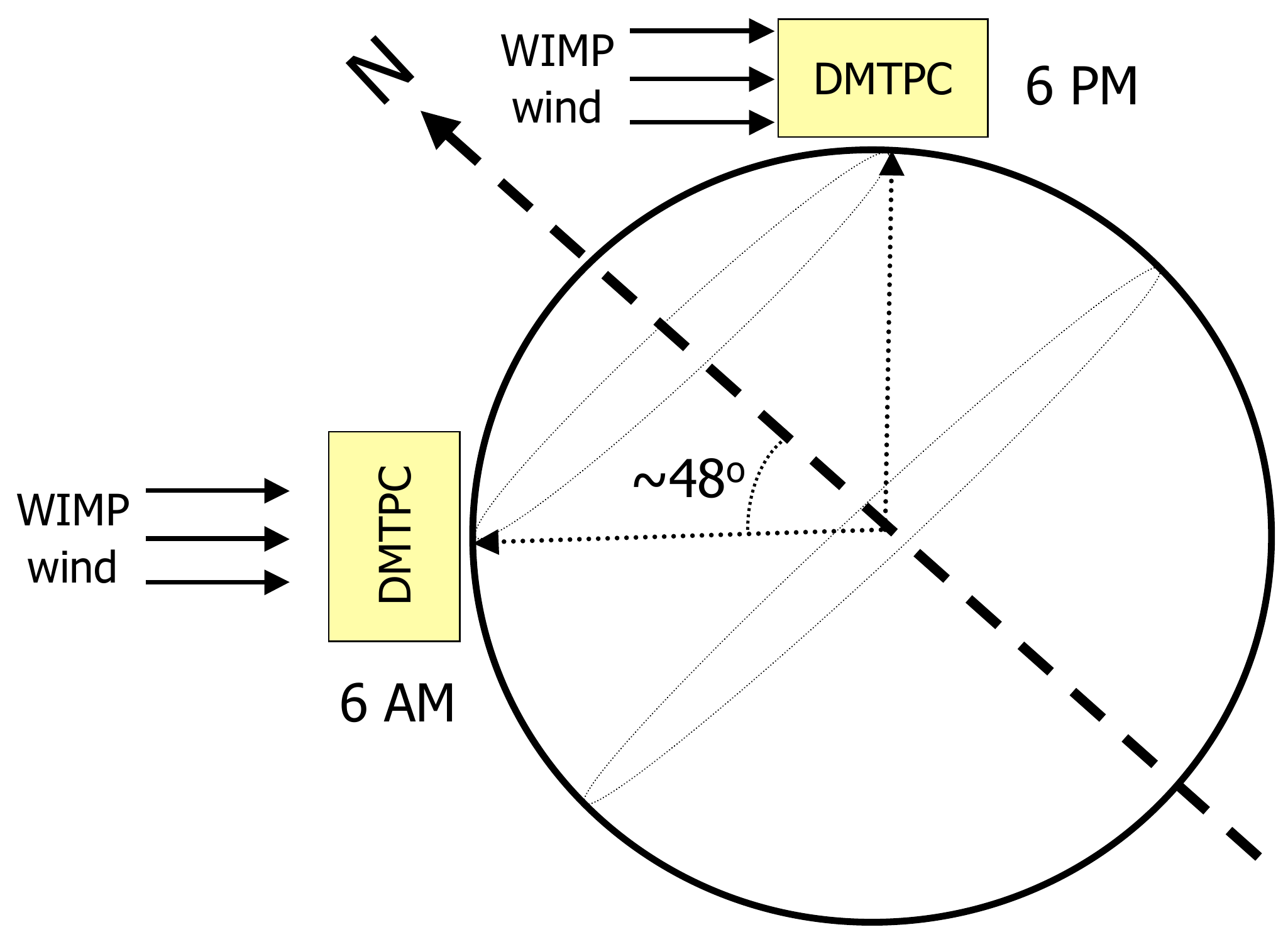}
\caption{ 
As the Earth spins  about its rotation axis, the average direction of the WIMP wind with respect to a DM detector 
 changes by nearly 90$^\circ$  every 12 sidereal hours.
} \label{dailyasymmetry}
\end{figure*}

A larger modulation of the WIMP signal 
was pointed out by Spergel\cite{spergel} in 1988 and is illustrated in Figure~\ref{dailyasymmetry}. The Earth spins around its axis with a period of 24 sidereal hours. Because its rotation axis is oriented at 42$^\circ$  with respect to the direction of the DM wind, an observer on Earth sees the average 
direction of the WIMPs change  
by nearly 96$^\circ$  every 12 sidereal hours.
 This modulation in arrival direction should be resolvable 
by a Dark Matter directional detector, e.g. a detector able to determine the direction of the DM wind. Most importantly, no known 
background is correlated with the direction of the DM wind. Therefore, a directional Dark Matter detector could hold the 
key to the unambiguous observation of WIMPs. 

In addition to background rejection, 
the determination of the direction of the arrival of Dark Matter particles can 
discriminate\cite{Alenazi,MorganGreenSpooner,CopiKrauss} 
between various DM halo distributions including the standard dark halo model,\cite{isothermal} 
models with streams of WIMPs, 
the Sikivie late-infall halo model,\cite{Sikivie} and other anisotropic models.  
The discrimination power improves if a determination of the vector direction 
of WIMPs is possible.    
This capability  makes directional detectors 
unique observatories for underground WIMP astronomy. 

\section{Requirements for a directional DM detector } 

\begin{figure*}[t]
\centering
\includegraphics[width=4in]{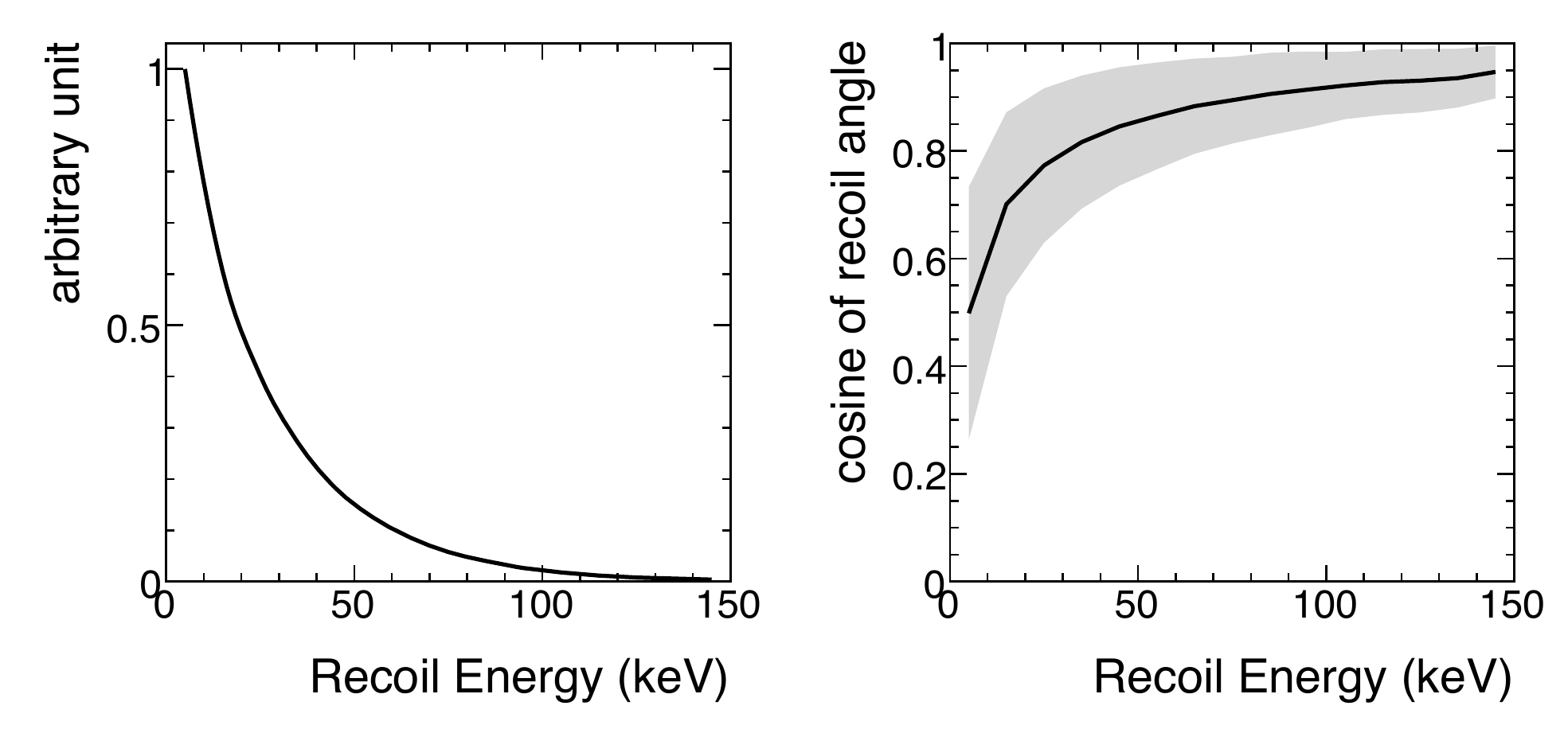} 
\caption{ 
Energy spectrum (left) and distribution of the cosine of the recoil angle (right) vs. recoil energy 
for fluorine  recoils induced by  100~GeV WIMPs  in CF$_4$. 
The recoil angle is defined as the angle between the direction of the 
recoil and the direction of the incoming particle. 
} \label{dmmc}
\end{figure*}

When Dark Matter interacts with normal matter it generates nuclear recoils with typical energies of a few tens of keV 
(Figure~\ref{dmmc}-left). The direction of the recoiling nucleus encodes the direction of the incoming DM particle 
(Figure~\ref{dmmc}-right). To  observe the daily modulation in the direction of the DM wind, 
an  angular resolution  of  20--30 degrees in the reconstruction of the recoil nucleus is required.

A key requirement for a directional detector is, therefore, to produce nuclear recoil tracks 
that are at least 1 mm in length, assuming sub-millimeter tracking resolution can be achieved. 
Long recoils are obtained by using a very dilute gas as a target material. 
As an example, a typical 50 keV fluorine recoil generated by an elastic collision of a 
WIMP in CF$_4$ at 40 torr produces a 2 mm track. 
The direction of the recoiling nucleus cannot be determined 
with a conventional DM detector using solid or liquid target because 
the typical length of a recoil track in such media 
is less than 1 $\mu$m,  well below the tracking resolution of the detector.

An ideal directional detector should  provide a 3-D reconstruction of the recoil track with a spatial resolution of a few hundred microns in all three coordinates. A 2-D reconstruction is still valuable, although it degrades the sensitivity.\cite{GreenMorgan} It is also very important to be able to determine the sense of the direction by 
 discriminating between the ``head'' and ``tail'' of the recoil track. The ability to reconstruct the sense as well as the direction of the recoil improves the sensitivity to DM by about one order of magnitude.\cite{GreenMorgan}

The advantages associated with directional detectors come at a price: a detector with a fiducial mass of a few tons, 
necessary to observe DM-induced nuclear recoils,
 would occupy thousands of cubic meters. It is, therefore, key to the success of the directional DM program to develop detectors 
with a low cost per unit volume. Since for standard gaseous detectors the largest expense is represented by the 
cost of the readout electronics, 
it follows that a low-cost read-out is essential to make DM directional detectors financially viable.

\section {Background suppression in directional detectors }

The main source of background for any DM experiment is due to electromagnetically interacting particles such as photons, 
electrons, and alpha particles. 
These particles are produced by natural radioactivity of the detector components and surrounding materials, 
as well as by cosmic rays. 
Most DM  experiments  can effectively suppress such backgrounds by using only radio-pure  materials 
in the fabrication of the apparatus, by constructing shielding around the detector, 
and by rejecting background events in the off-line analysis based on their characteristic signature.  

Directional detectors provide an additional signature that 
can be used to distinguish particle species: the 
correlation between the energy and length of the recoil track. 
As an example, in the case of CF$_4$ gas at a pressure of 50 torr, simulations show that a typical 30 keV fluorine 
recoil travels about 1 mm. 
A 15 keV electron, which produces the same net  ionization, travels about 30 mm and can be easily distinguished from the signal. 
A 7 keV $\alpha$ particle has a range of about 1 mm. 
However, since this energy is much smaller than that of a fluorine ion with the same range, such a low energy $\alpha$ particle 
can be distinguished from  a recoiling nucleus.

Neutrons constitute another source of background. 
While much less abundant than $\alpha$ particles, photons, or electrons, neutrons are  more insidious for DM searches because their signature is indistinguishable from  that of a WIMP. 
Neutrons arise from two independent processes: natural radioactivity (fission and ($\alpha,n$) reactions),  and interactions of high-energy muons in the rock surrounding the detector. The incidence of the former category, which is characterized by a low-energy spectrum of 
less than 10 MeV,\cite{MeiHime} can be effectively reduced with passive shielding (e.g. water or  polyethylene). 
Muon-induced neutrons have a much harder spectrum, with a substantial tail above 10 MeV\cite{MeiHime} which reduces the 
effectiveness of 
shielding. As a consequence, it is necessary to locate DM detectors in a deep-underground laboratory.  In addition, 
a neutron veto can  be used to improve the rejection of such backgrounds. 

Directional detectors have an additional advantage with respect to traditional DM detectors in distinguishing 
neutron backgrounds from WIMP signals: while the signal is expected to be correlated with the direction of the 
WIMP wind, recoils from neutrons are either isotropically distributed, or point back to some particular regions 
in the detector where radioactive material is present. 

Finally, directionality is key to distinguish between  coherent  scattering from solar neutrinos from WIMP signal: 
while neutrinos will point back to the Sun, the signal will follow the direction of the DM wind. 
 
\section {Existing directional detectors}
The properties of the current directional  Dark Matter experiments are summarized in Table~\ref{tableExp}.  
Most directional detectors are Time Projection Chambers\cite{TPC} (TPCs) that use the gas 
both as target and detector material. 
The pressure of the gas  inside the vessel is chosen in the range 40-100 torr.  
At this  pressure a typical collision of a 100 GeV WIMP of 220 km/s with a gas molecule causes a 
nucleus to recoil about 1-2 mm. 
The ionization electrons produced by the recoiling nucleus drift in the gas along 
a uniform  electric field toward an anode plane. The intense electric field around the anodes 
triggers an avalanche that amplifies the signal
which  allows us to reconstruct the projection of the 3-D nuclear recoil on the 2-D anode plane. 
The  time between the beginning and the end of the signal determines the 
length of the nuclear recoil in the direction parallel to the drift. 
Because the energy loss (dE/dx) is not uniform along the trajectory, the sense of the recoiling nucleus 
can be determined   (``head-tail'' measurement). 

\begin{table}
\tbl{
Summary of existing directional Dark Matter experiments. 
}
{\begin{tabular}{@{}cccccc@{}} \toprule
Experiment  &  Gas     &  Interactions      & Technology        & Readout     \\
\colrule
DRIFT	    &   CS$_2$ &  SI                &  Negative ion TPC & MWPC \\
NewAge	    &   CF$_4$ &  SD/SI             &  TPC              &  $\mu$PIC \\
MIMAC	    &   $^3$He/CF$_4$    &  SD/SI   &  TPC               & Electronic 2-D  \\
DMTPC	    &   CF$_4$           &  SD/SI   &  TPC               & Optical 2-D    \\
\colrule
\end{tabular}}
\label{tableExp}
\end{table}

The field of Dark Matter directional detection was pioneered by the DRIFT\cite{DRIFT} 
experiment, which uses a negative ion time projection chamber\cite{NITPC} filled with 
carbon-disulfate (CS$_2$) gas at a pressure of 40 torr.  
Drifting negative ions instead of electrons reduces diffusion,  
thereby allowing for  drift distances 
of up to 50 cm without significant loss in spatial resolution
and without the need for a magnetic field. 
The detector is read out by multi-wire proportional counters (MWPCs). 

A full-size DRIFT module\cite{DRIFT2a} is currently taking data at the Boulby Underground Laboratory at a depth of 
2,805 meters water equivalent. 
The fiducial volume of this detector is $1\times 1\times 1$ m$^3$, corresponding to 167 g of CS$_2$.  
The apparatus consists of a 1.5 m$^3$ low-background stainless steel vacuum vessel containing two back-to-back TPCs with a shared, vertical, central cathode constructed of 20 $\mu$m stainless steel wires. 
Two field cages, located on either side of the central cathode, define  two 50 cm long drift regions.  
Charge readout of tracks was provided by two MWPCs each comprised of an anode plane of 20 $\mu$m stainless steel wires 
sandwiched between two perpendicular grid planes of 100 $\mu$m stainless steel wires. 
The detector is operated at a gas gain of 10$^3$, and the drift electric field is 624 V/cm.
Both cathode and anode wires are multiplexed and reduced to 8 channels of readout per wire plane. 
Because the pitch of the wires in the  MWPC is 2 mm, 
the eight adjacent readout lines (either grid or anode) sample a distance of 16 mm, 
which is  sufficient to fully contain a  WIMP candidate. 

The gas gain of the detector is monitored every 6 hours with a precision of 2\%
by using  a retractable $^{55}$Fe calibration source. 
The gamma-ray rejection capability of the detector, determined using $^{60}$Co sources, is  better than $8\times10^{-6}$. 
A $^{252}$Cf source is used to demonstrate an efficiency of 94\% in detecting nuclear recoils of sulfur (carbon) above a threshold of 47 (31) keV. The efficiency decreases to 60\% after off-line WIMP selection criteria are applied.

While analyzing the 10.2 kg days of data collected underground with a shielded DRIFT detector, 
an unexpected population of nuclear recoil events has been observed, and shown to be due to the decay of the daughter nuclei of $^{222}$Rn which have attached to the central cathode. Reduction of the radon inside the detector components, as well as development of additional measures to remove or veto Rn events, is underway. 

Laboratory runs using $^{252}$Cf sources demonstrate that  DRIFT  can measure the range of the recoil tracks in 3-D with 
adequate precision,\cite{DRIFT-RANGE} and show statistical discrimination in determining the sense of the recoil 
down to a threshold of 50 keV.\cite{DRIFT-HT} 
To improve the  3-D reconstruction capability of the detector, 
 the possibility of using Micromegas\cite{micromega} detectors as an alternative form of 
readout\cite{micromegaDM} is being investigated.  

The  NewAge\cite{NEWAGE,NEWAGE2} experiment 
uses CF$_4$ gas to detect spin-dependent interactions of WIMPs.
The apparatus is a time projection chamber with a micro pixel ($\mu$-PIC) readout. 
A $\mu$-PIC is a two-dimensional position-sensitive detector manufactured with printed circuit board technology, and provides gas multiplication and readout. 
Additional gas amplification\cite{Miuchi2007}  is provided by a gas electron multiplier (GEM), for a total gas gain 
of $5\times 10^4$.

A NewAge detector with a fiducial volume of 23$\times$28$\times$30 cm$^3$ is currently taking data in the Kamioka mine. 
The detector is contained in a stainless-steel vacuum chamber and operates at a pressure between 30 and 150 torr. 
The drift length is limited to 30 cm by transverse  diffusion. 
The uniformity of the electric field in the drift region is maintained by a field-shaping pattern on a fluorocarbon circuit board surrounding the detection volume.  
An $\alpha$ source  is used to calibrate and monitor the gain of the detector. The energy resolution is measured to be better than 70\% in the  range relevant to DM studies, while 
the spatial resolution is 800 $\mu$m. 
The efficiency of the detection of nuclear recoils is measured to be 40\% at 100 keV using a $^{252}$Cf source. 
The gamma ray rejection factor is measured using $^{137}$Cs to be $6\times 10^{-4}$.

Spin-dependent  interactions are also pursued by the MIMAC\cite{MIMAC} collaboration. MIMAC is a micro-TPC with an avalanche amplification with a pixelized anode. The target materials are  $^3$He and CF$_4$. The use of $^3$He is motivated 
by its light mass, which makes it more sensitive to low mass WIMPs. In addition, $^3$He has unique properties in terms of gamma ray rejection power and discrimination against neutron backgrounds.\cite{MIMAC}
MIMAC plans to run at two pressures: high pressure (1-3 bar) in the initial phase of the experiment, 
and low pressure (100-200 mbar) after DM candidates will be observed and  directional information will be needed 
to discriminate against backgrounds. 

Finally, a group from Nagoya University has proposed\cite{emulsions}
to use emulsions to image the  nuclear recoils caused by WIMP interactions. 
Emulsions, with their sub-micron 3-D spatial resolution, can determine the direction of WIMPs even from very short recoils. 
The target density can, therefore, be much higher than in gaseous directional detectors, 
and a large fiducial mass can be contained in a relatively small volume.  
The limitation of this detector has to do with the fact that 
long integration time may smear the directional signal. In addition, 
 an efficient rejection of backgrounds such as very low energy $\alpha$ 
particles and electrons may be more challenging to achieve than in other directional detectors.

\section{The DMTPC detector }  

\begin{figure*}[t]
\centering
\includegraphics[width=8cm]{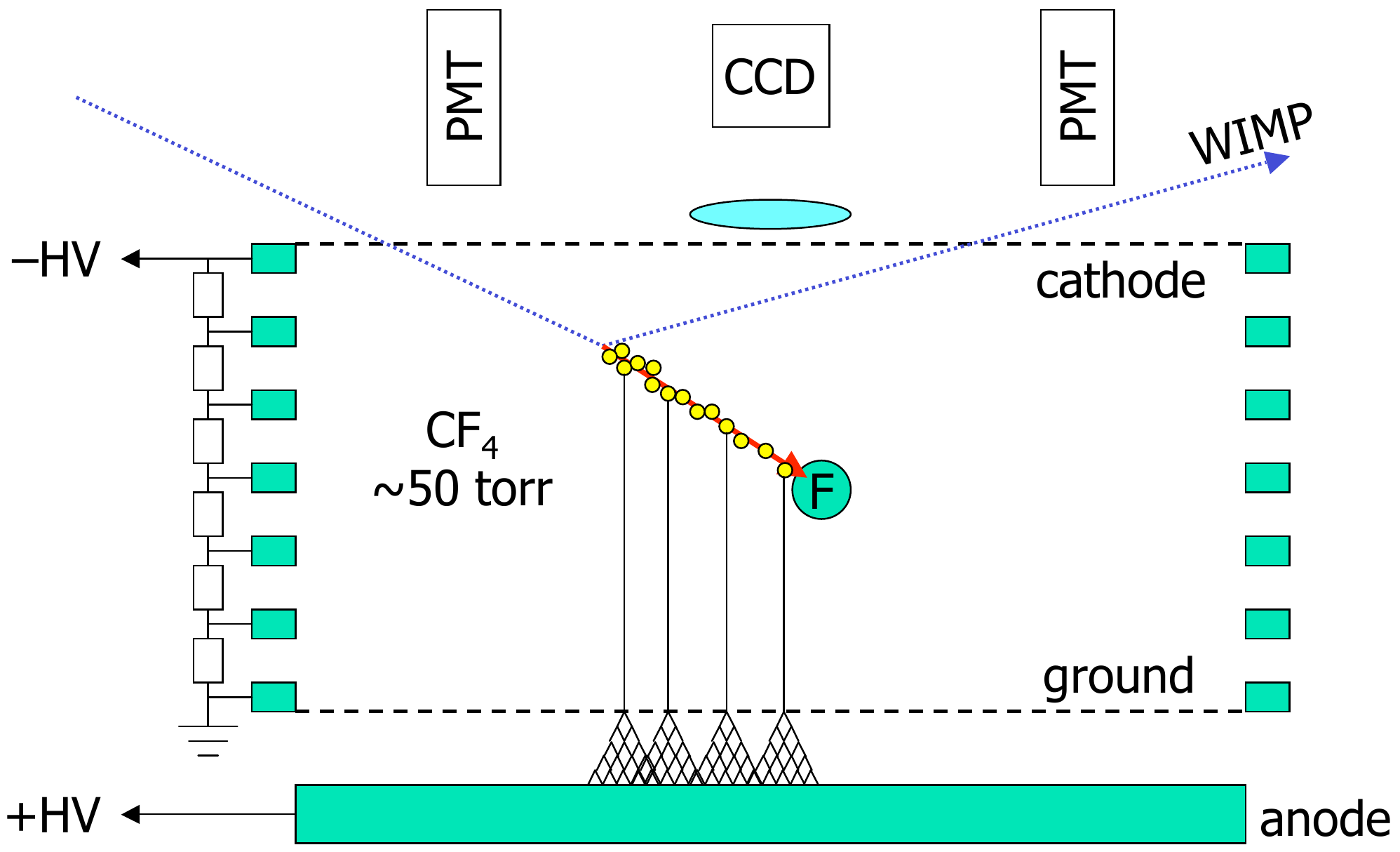}
\caption{ Illustration of the DMTPC detector concept. 
} \label{detectorconcept}
\end{figure*}

The DMTPC detector is a low-pressure TPC with optical readout. 
The detection principle is illustrated in Figure~\ref{detectorconcept}. 
The TPC is filled with tetrafluoromethane (CF$_4$) at a pressure of about 50 torr. 
At this pressure, a typical 50 keV nuclear recoil has a length of about 2 mm. 
The  electrons liberated by the recoiling nuclei drift toward the amplification region, where 
photons are produced together with electrons in the avalanche process. 
A CCD camera mounted above the cathode mesh images these photons  recording  
the projection of the 3-D nuclear recoil on the 2-D amplification plane. 
An array of photomultipliers (PMTs) mounted above the cathode mesh 
measures the length of the recoil in the drift direction.

\subsection{$CF_4$ gas}   
 CF$_4$ was chosen as  target material primarily because of its high content in fluorine. Fluorine is an
excellent element\cite{ellis} to detect spin-dependent interactions on protons, 
due to its large spin factor and isotopic abundance, as shown in  
Table~\ref{tableF}. 

In addition, CF$_4$ 
is an excellent detector material\cite{Pansky,Christophorou}. Its 
good scintillation properties are very important because of the  
optical readout. Recent measurements\cite{asher} indicates that in CF$_4$  
the number of photons produced 
between 200 and 800 nm 
is about 1 for every 3 avalanche electron. 
Moreover, the low transverse diffusion characteristic of CF$_4$ allows for a good spatial 
resolution in the reconstruction of the recoil track despite the long (25 cm) drift distance. 
Finally, CF$_4$ is non-flammable and non-toxic, and, therefore, safe to operate underground. 

\begin{table}
\tbl{Spin of the nucleus (J), nuclear spin factor ($\lambda^2$J(J+1)) and 
abundance in nature of various isotopes considered for 
SD-interaction measurements. The figure of merit, to be used to 
compare the various isotopes, is defined as the product of the square of the nucleus mass, 
the number of isotopes per kg, and the  spin factor. }
{\begin{tabular}{@{}ccccc@{}} \toprule
Isotope  &  J    &  $\lambda^2$J(J+1)  & Natural abundance & Figure of merit\\
\colrule
$^{1}$H	 & 1/2   &  	0.75 &  100\% & 	5      \\ 
$^{19}$F  & 1/2  &  	0.65 &  100\% & 	74     \\
$^{23}$Na & 3/2  &  	0.04 &  100\% & 	6      \\
$^{73}$Ge & 9/2  &  	0.06 &  7.8\% & 	2      \\
$^{93}$Nb & 9/2  &  	0.16 &  100\% & 	91     \\
$^{127}$I & 5/2  &  	0.01 &  100\% & 	5      \\
$^{129}$Xe & 1/2  &  	0.12 &  26\% & 	25             \\
\colrule
\end{tabular}}
\label{tableF}
\end{table}

\subsection{Amplification region}   
\begin{figure*}[t]
\centering
\includegraphics[width=\textwidth]{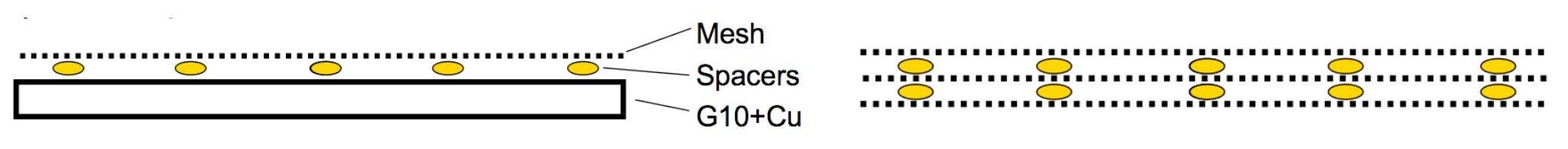}
\caption{ 
Two designs for the DMTPC amplification region: the ``mesh-plate'' design is shown on the left, and the triple mesh design  is shown on the right.   
} \label{amplification}
\end{figure*}

Two alternative implementations of the amplification region\cite{meshpaper}  are 
illustrated in Figure~\ref{amplification}.  
In the first design (Figure~\ref{amplification}-left), the amplification is obtained by 
applying  a large potential difference 
($\Delta$V = 0.6--1.1 kV) 
between a copper plate and a conductive woven mesh. A uniform distance 
of 540 $\mu$m between the plate and the mesh 
is ensured by the use of fishing wires spaced 2 cm apart. 
The copper or stainless steel mesh is made of 28 $\mu$m wire with a pitch of 256 $\mu$m. 
The pitch of the mesh determines the intrinsic spatial resolution of the detector. 

In a second design (Figure~\ref{amplification}-right), 
the copper plate is replaced with two additional woven meshes.  
This design has the advantage of creating a transparent amplification region, 
which enables a single CCD camera to image two drift regions. 
This allows for a substantial cost reduction.

\subsection{Optical readout}   
Optical readouts for directional DM experiments were first explored in the 1990s\cite{sandiego}. 
CCDs provide a true 2-D readout at
a much lower cost per channel than any other technology used in particle physics. 
A modern low-noise CCD camera with 4 megapixels can be purchased today for a few thousand US dollars, 
which corresponds to $10^{-3}$ dollars/channel. 
Because the cost of the readout electronics dominates the overall cost of a directional DM detector, the choice of an optical readout makes directional detectors economically viable. 
Widespread use of CCD cameras from cell phones to medical x-ray drives the development of new and more efficient CCDs. As a result, better products are available every year at decreasing prices. 

A detector with an optical readout needs a well-designed optics. 
Because the average ionization energy in CF$_4$ is 54 eV,\cite{NEWAGE2} 
a typical nuclear recoil with an energy of 50 keV will produce about $10^3$ primary electrons. 
Typical gas gains in the amplification region are of the order of $10^5$, and the ratio of photons to electrons produced in the amplification is 1:3. 
Therefore, the number of photons produced in the amplification region is about $10^7$. 
The fraction of the photons that are collected by the lens and detected by the CCD 
depends on the solid angle covered by the lens.
As an example, for a lens aperture of 25 mm and a distance of 30 cm from the amplification plane, the number of photons detected is $10^2$--$10^3$, 
after taking into account the transmittance of the meshes
and the lens, and the quantum efficiency of the CCD. 
Another consideration is 
the area of the amplification region that can be covered by a single camera.
Optics with wide a field of view reduces the number of cameras and, therefore, the cost
of the detector.
Optimizing the lens design is clearly of crucial importance.

\subsection{Measurement strategy}   

The DMTPC detector is designed to measure the following quantities: 
\begin{romanlist}[(iii)]
\item{} the number of photons observed in the CCD camera; 
\item{} the 2-D image of the recoiling nucleus; 
\item{} the distribution of energy loss along the recoil track; 
\item{} the width and integral of the PMT signal;
\item{} the electronic signal produced on the amplification plane.   
\end{romanlist}

The energy of the nuclear recoil is independently determined  
by  the measurement of the number of photons observed in the CCDs, 
the integral of the electronic signal produced on the amplification mesh, 
and the integral of the PMT signal. The redundancy in the design is intentional 
and has the goal of maximizing the robustness of the measurement. 
The track length and direction of the recoiling nucleus is reconstructed 
by combining the measurement of the projection along the amplification 
plane (from pattern recognition in the CCD) and the projection along the 
direction of drift, determined from the width of the signal recorded in the PMTs. 
The sense of the recoil track can be determined by the Bragg curve,
the characteristic variation of the energy deposit along the length of the track. 

The CCD images have a long (about 1 second) exposure. 
If during the exposure a trigger is generated by the PMT or the 
electronic read out of the amplification plane, the CCD is read out and the event is saved to disk. 
Otherwise, the CCD is reset, to minimize dead time. 

The combination of the measurements described above is very effective in reconstructing the energy, direction, 
and sense of  nuclear recoils from WIMPs. 
In addition,  an excellent rejection of the electromagnetic backgrounds can be obtained 
by combining the measurement of the energy and length of the recoil. 
The gamma ray rejection factor, measured using a $^{137}$Cs source, is better than 2 parts per million.

\subsection{Detector design} 

\begin{figure*}[t]
\centering
\includegraphics[width=120mm]{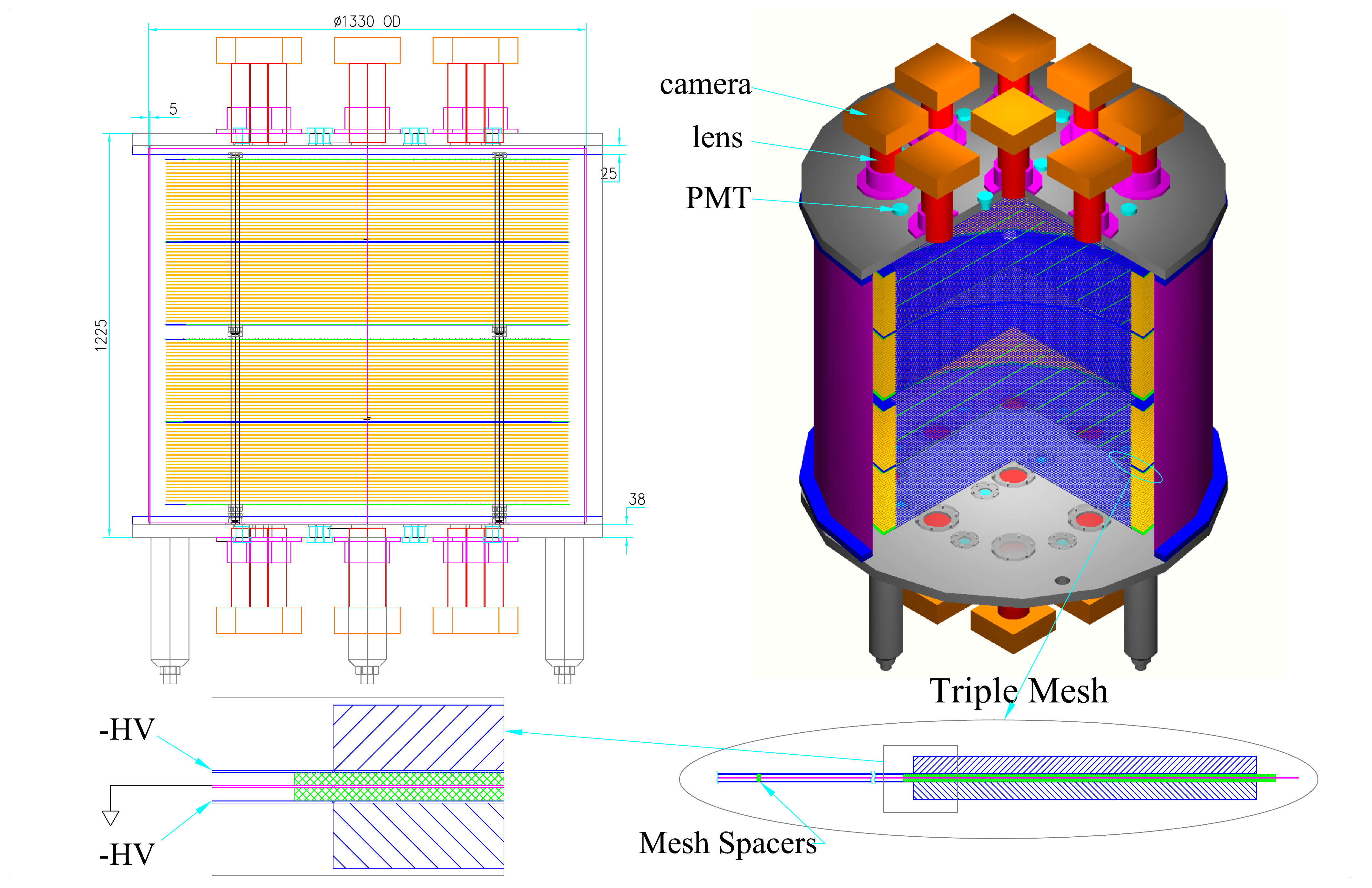}
\caption{Preliminary drawings of the 1 $m^3$ DMTPC detector. } 
\label{DMTPC1m}
\end{figure*}

The preliminary design of a 1-m$^3$ DMTPC detector is shown in Figure~\ref{DMTPC1m}. 
The apparatus consists of a stainless steel vessel of 1.3 m diameter and 1.2 m height.
Nine CCD cameras and nine PMTs are mounted on each of the top and bottom plates of the vessel, separated from the active volume of the detector by an acrylic window.  The detector consists of two optically separated regions. Each of these regions is equipped with a triple-mesh amplification device, mounted in between two symmetric drift regions.  Each drift region has a diameter of 1.2 m and a height of 25 cm, for a total active volume of 1 $m^3$. 
A field cage made of stainless steel rings keeps the uniformity of the electric field within 1\% in the fiducial volume. A gas system recirculates and purifies the CF$_4$. 

All materials used inside the active volume of the detectors are required to be radio-pure 
to limit backgrounds from internal radioactivity. Pure copper, stainless steel, and acrylic  
are known to satisfy these requirements.\cite{exopaper} 
Because all CCD cameras and lenses are outside the active volume, their contribution to  internal radioactivity is minimal. 

When operating the detector at a pressure of 50 torr and 21 degrees C, this module will contain 250 g of CF$_4$. 
Assuming an overall data-taking efficiency of 50\%, a one-year underground run will yield an exposure of 
45 kg-days. 

\subsection{ R\&D results} 


The current DMTPC prototype 
consists of two optically independent regions contained in one stainless steel vessel. Each region is a cylinder with 25 cm diameter and 25 cm height contained inside a field cage. 
The amplification is obtained by using a mesh-plate design. The detector is read out by two CCD cameras, each imaging one drift region. The optical system uses two Nikon photographic lenses with 
f-number of 1.2 
and 
focal length of 55 mm, 
and two Apogee U6 CCD cameras\cite{apogee} equipped with  Kodak 1001E CCD chips. Because the total area imaged is $16\times16$~cm$^2$, the detector has an active volume of about 10 liters.

\begin{figure*}[t]
\centering
\includegraphics[width=\textwidth]{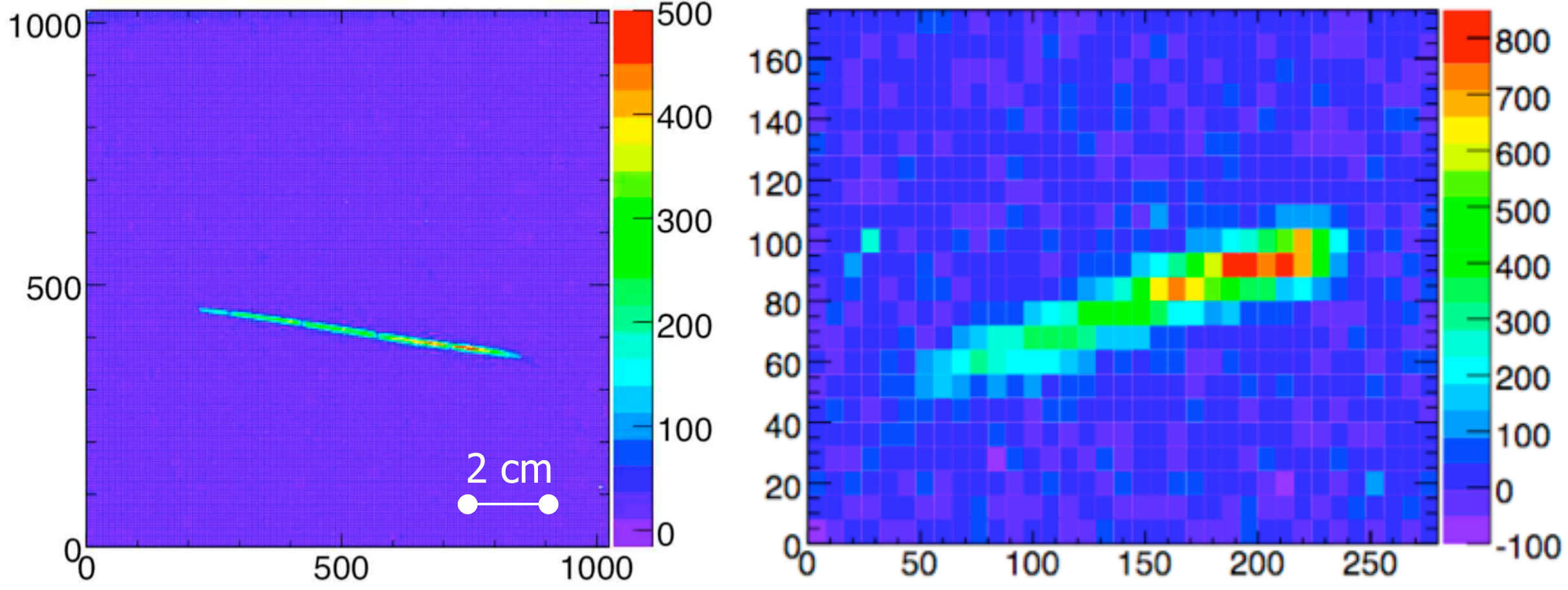}
\caption[]{
Left: an $\alpha$ track stopped in  CF$_4$ at a pressure of 100~torr. 
 The particle travels left to right. The red region on the right of 
the track corresponds to the Bragg peak. 
Right: a nuclear recoil from a low-energy neutron 
in  CF$_4$ at 75~torr. 
The neutron, produced by a $^{252}$Cf source, was traveling 
right to left. The higher $dE/dx$ visible on the right of the track 
is consistent with observation of the ``head-tail'' effect. 
} \label{events}
\end{figure*}

The main calibration source is a $^{241}$Am source that produces 5.5 MeV alpha particles. 
The image of one alpha particle stopped inside CF$_4$ at 100~torr is shown in Figure~\ref{events}(left). 
This source is used to study the gain of the detector as a function of the voltage 
in the amplification region and gas pressure, while a $^{55}Fe$ is used for absolute 
energy calibration. 
The alpha source is also used to measure the resolution as a function of the drift distance of the 
primary electrons to quantify the effect of the diffusion. These studies\cite{headtailpaper} 
show that the transverse diffusion is $\approx$ 1 mm for a drift distance of 25 cm.

The performance of the DMTPC detector in determining the sense and direction of nuclear recoils has been evaluated by studying the recoil of fluorine nuclei in interaction with low-energy neutrons. 
When the  nuclear recoil produced is below 1 MeV, the energy deposition  decreases along the path of the recoil, allowing for the identification of  the ``head'' (``tail'') of the event by a smaller (larger) energy deposition. 
To quantify this effect, we define the skewness
$s \equiv \langle x^3 - \langle x \rangle^3 \rangle /
          \langle x^2 - \langle x \rangle^2 \rangle^{3/2}$
where $x$ is the position along the recoil track,
signed according to the direction of the neutron beam,
and the angle brackets denote averages weighted by the light yield.
Decreasing light yield along the recoil track should result in a negative value of $s$.

The initial measurements, performed with a wire-based prototype, used 14 MeV neutrons from a deuteron-triton generator. 
The reconstructed recoils had energy between 200 and 800 keV. 
The ``head-tail'' effect was  observed with a significance of  8 $\sigma$.\cite{headtailpaper}

\begin{figure}
\center
\includegraphics[width=0.5\textwidth]{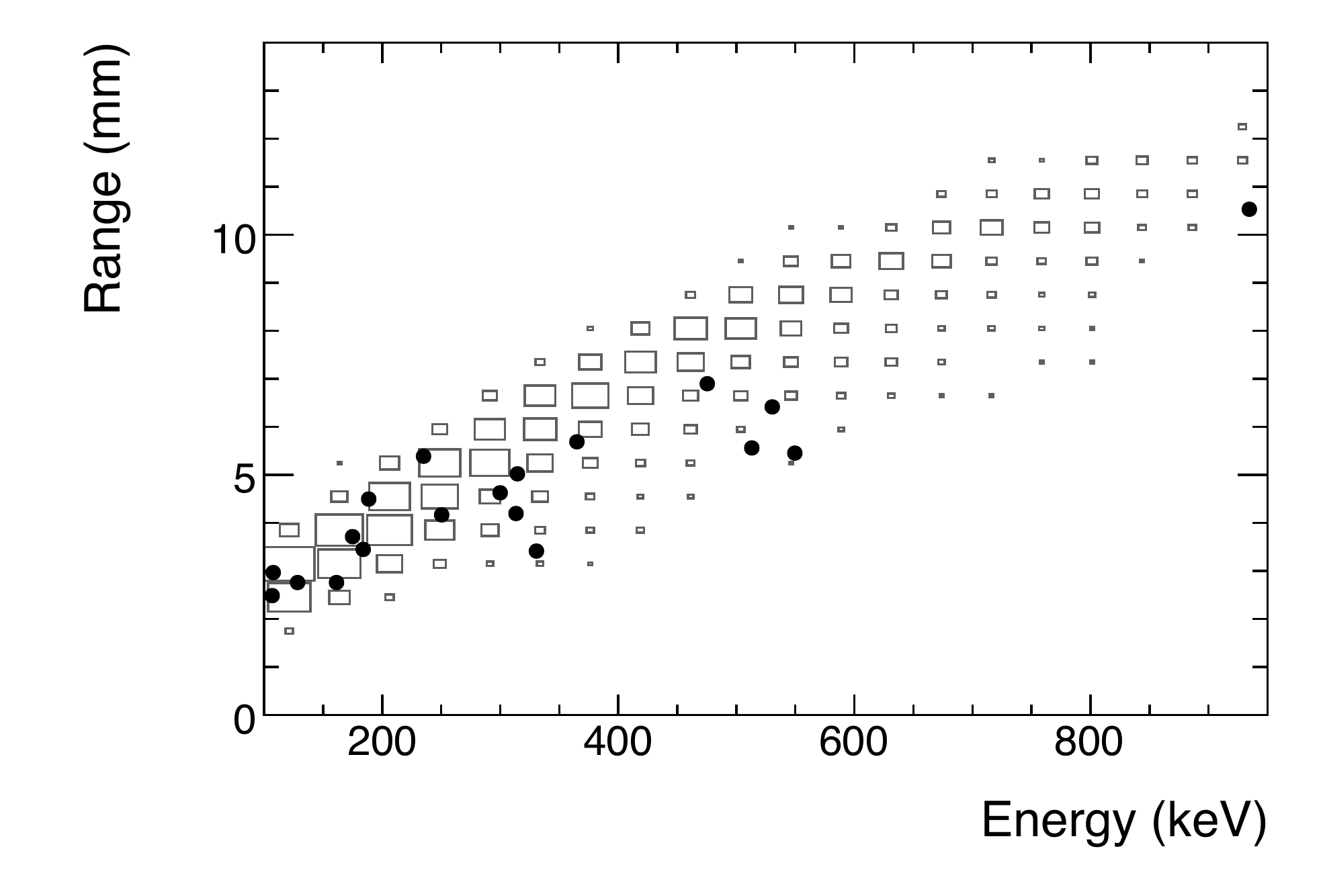}%
\includegraphics[width=0.5\textwidth]{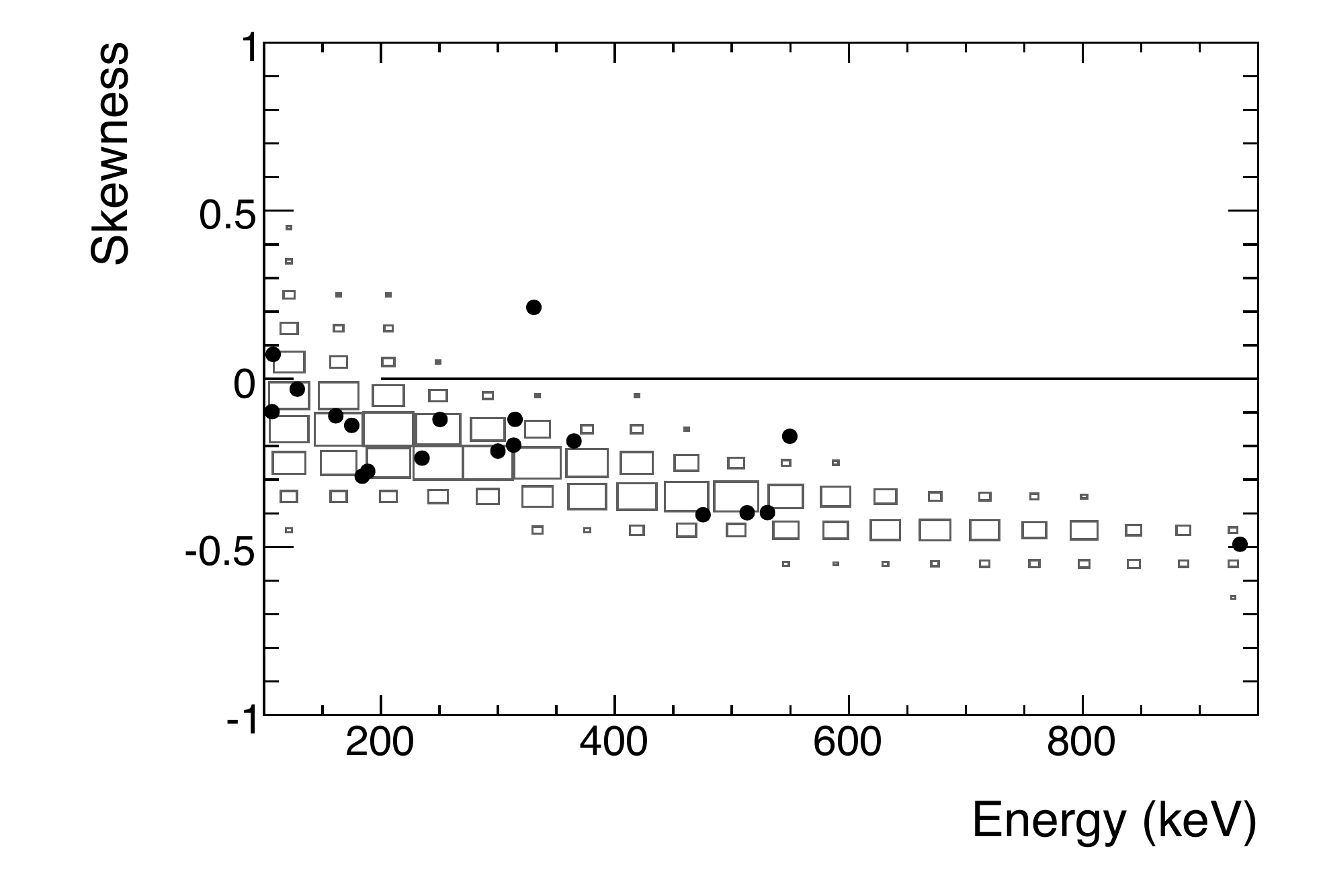}
\caption{Range (left) and skewness (right) vs. reconstructed energy 
for nuclear recoil candidates in a $^{252}$Cf exposure at 75~torr. 
Black points are data, the box-histogram is simulation.
\label{CfRun}}
\end{figure}

Subsequent measurements used lower energy neutrons generated by a 
$^{252}$Cf  source, and a  mesh-based detector to obtain a  
2-D reconstruction of the nuclear recoils.   
Better sensitivity  to lower energy thresholds was achieved by lowering the CF$_4$ pressure to 75 torr. 
 Figure~\ref{events}(right) shows a Cf-induced nuclear recoil reconstructed in the DMTPC detector.  
The neutron was traveling right to left.  The decreasing $dE/dx$ along the track direction clearly visible in the image  
proves that the detector is able to determine the sense of the direction on an event-by-event basis. 
Measurements of the length  and skewness of the recoil tracks as a function of their  energy   
are shown in Figure~\ref{CfRun}. The data is in good agreement with the predictions of the SRIM\cite{SRIM} MC. 
These measurements establish\cite{meshpaper} good head-tail discrimination for recoils above 100 keV. 
The ``head-tail'' discrimination is expected to extend to recoils above 50 keV 
when the  detector is operated at a pressure of 50 torr. 

%

\subsection{Expected sensitivity}

The sensitivity of the DMTPC detector to SD interactions of WIMPs on protons has been studied. 
In the calculation, it is assumed that the detector will be operated at a threshold of 50 keV   
in an underground laboratory at a depth of 4,300 m.w.e. and that 
it will be surrounded by a neutron shielding consisting of 40 cm of polyethylene. 
No significant internal backgrounds are assumed to be present above threshold. 

\begin{figure}[thb]
\center
\includegraphics[width=9cm]{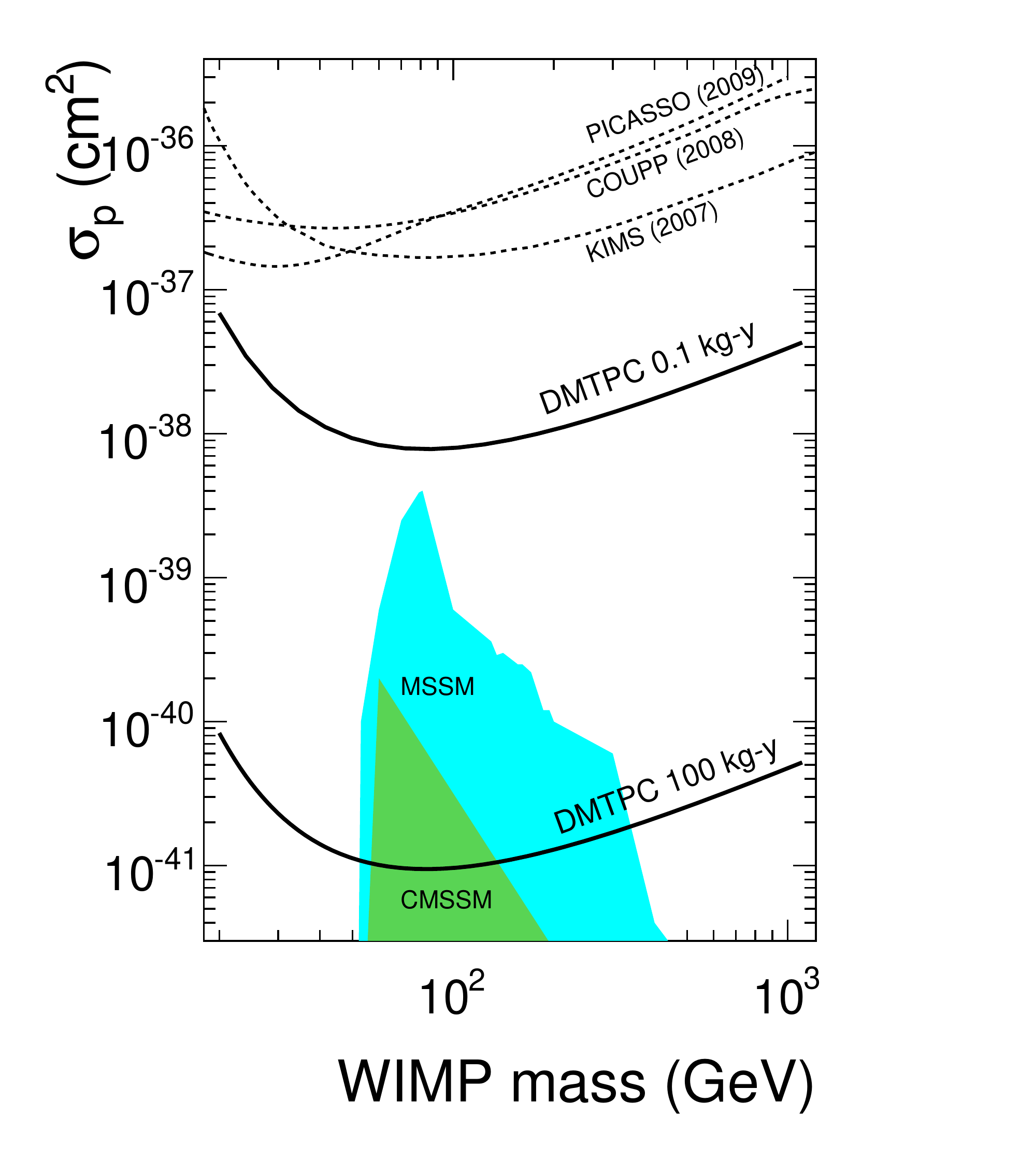} 
\caption{ 
Expected sensitivity (90\% C.L.) to spin-dependent WIMP scattering
on protons for a DMTPC detector assuming two different exposures.  
The dashed lines show the best published limits. 
The shaded area shows the allowed region from MSSM calculations.   
\label{sensitivity}}
\end{figure}

The limits expected for exposures of 0.1 and 100 kg-years  are shown in Figure~\ref{sensitivity}. 
The dashed lines in the same figure show the present best limit on spin-dependent interactions from the 
KIMS,\cite{KIMS2007} 
COUPP,\cite{COUPP2008} and NAIAD\cite{NAIAD2005}   collaborations. 

Improvements of a factor of 50 over the best existing measurements can be obtained 
with 0.1 kg-year exposure, which can be achieved operating a 1-m$^3$  detector for less than one year.
The high sensitivity is achieved despite of 
the limited target mass due to the excellent properties of fluorine for the study 
of SD interactions (Table~\ref{tableF}) and because of the excellent 
background rejection capability of the detector. 
 
A larger detector, with an active mass of $10^2$--$10^3$ kg, will be able to explore 
a significant portion of the Minimal Supersymmetric Standard Model (MSSM) parameter space\cite{ellisSD}. 
This detector is an ideal candidate for the DUSEL laboratory in South Dakota.

\section{ Conclusion } 
Directional  detectors can  provide an unambiguous positive 
observation of Dark Matter particles even in presence of 
insidious backgrounds, such as neutrons or neutrinos. 
Moreover, the measurement of the direction of the incoming WIMPs 
will allow us to discriminate between the various DM models in our galaxy, 
opening the path to underground WIMP astronomy.  

In the past few years, several groups have investigated new ideas to develop directional 
Dark Matter detectors. Low-pressure TPCs are well suited for the purpose, if a 
3-D reconstruction of the nuclear recoil can be achieved using an inexpensive readout,  
and an effective rejection of electromagnetic backgrounds can be obtained. 

The DMTPC experiment addresses both of these requirements.  
This detector measures the energy, direction, and sense of the nuclear recoils produced in elastic collisions 
of WIMPs in low-energy CF$_4$ gas.  The redundancy between the various measurements allows for excellent background 
rejection, while the use of an optical readout substantially reduces the costs, which makes a ton-size directional 
detector economically viable. The choice of  CF$_4$  as the target material makes this detector particularly suited 
to study SD interactions, complementing the effort of most other DM experiments (e.g. noble liquid detectors),  
more suitable for the study of SI interactions. 

\section*{Acknowledgments}
I am grateful to my DMTPC collaborators and, in particular,  
to D.~Dujmic, H.~Wellenstein, and A.~Lee for contributing some of the 
plots and results used in this paper. 
In addition, I wish to thank M.~Morii and J.~Battat for 
useful discussions and for proofreading the manuscript.  
This work is supported by the  U.S. Department of Energy (contract number DE-FG02-05ER41360) 
and the MIT Physics Department.

\end{document}